\documentclass[a4paper]{jpconf}
\usepackage{graphicx}

\newcommand{\href}[2]{ \, #2}

\newcommand{\comment}[1]{}

\newcommand{\lr}[1]{ \left( #1 \right) }
\newcommand{\lrs}[1]{ \left[ #1 \right] }

\newcommand{\vev}[1]{ \langle \, #1 \, \rangle }

\begin{document}

\title{Axial Magnetic Effect and Chiral Vortical Effect with free lattice chiral fermions}

\author{P. V. Buividovich}

\address{Institute for Theoretical Physics, Regensburg University, Universit\"{a}tsstrasse 31, D-93053 Regensburg, Germany}
\ead{pavel.buividovich@physik.uni-regensburg.de}

\begin{abstract}
 Following the recent work by Braguta \emph{et al.}, we perform an indirect calculation of the chiral vortical conductivity $\sigma_V$ with free lattice overlap fermions by measuring the response of the energy flow to the constant axial magnetic field. We find that the measurements are contaminated by very large finite-volume artifacts. However, the continuum result for finite-temperature contribution to the chiral vortical conductivity $\sigma_V = \frac{T^2}{12}$ is still reproduced in the limit of infinite volume at fixed temperature.
\end{abstract}

 Transport phenomena in chirally imbalanced medium have become a subject of intense research in recent years. These phenomena are closely related to quantum anomalies and for that reason are extremely interesting from theoretical point of view. It has been also argued that they might manifest themselves in heavy-ion collision experiments (see \cite{Kharzeev:12:1} for an extensive collection of reviews). Very recently, these arguments have been supported by fully nonlinear simulations of the anomalous hydrodynamics \cite{Hongo:13:1}.

 One of such anomalous transport phenomena is the Chiral Vortical Effect (CVE) - the generation of axial current $j_{A \, i}$ (where $i = x, y, z$ labels spatial coordinates) in the direction of the medium vorticity $\omega_i = \epsilon_{i j k} \partial_j v_k$, where $v_i$ is the medium velocity \cite{Erdmenger:08:1, Surowka:08:1, Landsteiner:11:1}:
\begin{eqnarray}
\label{CVEClassical}
 j_{A \, i} = \sigma_V \, \omega_i ,
\\
\label{ChiralVorticalConductivity}
 \sigma_V = \lr{\frac{\mu_A^2 + \mu_V^2}{4 \pi^2} + \frac{T^2}{12}}
\end{eqnarray}
where $\sigma_V$ is the so-called Chiral Vortical Conductivity, $\mu_A$ and $\mu_V$ are the chiral and the baryon chemical potentials and $T$ is the medium temperature. In other words, vortices in a chiral liquid create chirality imbalance. Interestingly, this effect exists even in a neutral medium at finite temperature. It also turns out that the chiral vortical conductivity (\ref{ChiralVorticalConductivity}) might receive nontrivial contributions due to gauge interactions between fermions \cite{Son:12:1, Ren:12:1}.

 In the linear response approximation, the chiral vortical conductivity $\sigma_V$ is expressed in terms of the static correlator of the axial current density $j_{A \, i}$ and the energy flow vector $T_{0 i}$ \cite{Landsteiner:11:2}:
\begin{eqnarray}
\label{CVELinearResponse}
 \sigma_V = \lim\limits_{p_k \rightarrow 0} \frac{i \, \epsilon_{i j k}}{2 p_k} \int dt d^3x \vev{j_{A \, i}\lr{x, t} T_{0 j}\lr{0, 0} } \, e^{i p_k x_k}  ,
\end{eqnarray}
where
\begin{eqnarray}
\label{EnergyMomentumTensor}
 T^{0 i} = \frac{i}{2} \bar{\psi} \gamma_0 \partial_i \psi +  \frac{i}{2} \bar{\psi} \gamma_i \partial_0 \psi
\end{eqnarray}
are the components of the fermionic energy-momentum tensor which are responsible for the energy flow in the $i$-th direction.

 While it is conceptually very difficult to implement matter rotation on the lattice mainly due to problems with periodic boundary conditions (see \cite{Yamamoto:13:1} for some work in this direction), the correlator (\ref{CVELinearResponse}) is quite easy to measure on the lattice. One can simplify the problem even further by noticing that the expression (\ref{CVELinearResponse}) in fact also describes another anomalous transport phenomenon (sometimes termed Axial Magnetic Effect) - the generation of nonzero energy flow $T_{0 i}$ in the direction of the axial magnetic field $B_{A \, i} = \epsilon_{i j k} \partial_j A_k$. Here $A_k$ is the axial gauge field which couples to Dirac fermions as $A_{\mu} \bar{\psi} \gamma_5 \gamma_{\mu} \psi$. In close analogy with the lattice studies of the Chiral Magnetic Effect (see e.g. \cite{Yamamoto:11:1}) one can then introduce a constant axial magnetic field on the lattice and measure the energy flow along it. Technically, an advantage of the approach of \cite{Braguta:13:1} is that only the expectation value of a single fermionic bilinear operator should be measured, rather than the correlator of two operators in (\ref{CVELinearResponse}). This significantly simplifies algorithms and improves the signal-to-noise ratio. The results of such measurements in $SU\lr{2}$ quenched lattice gauge theory with zero chiral and baryonic chemical potentials were reported recently in \cite{Braguta:13:1}. It was found that the chiral vortical conductivity is equal to zero within statistical errors in the confinement phase and is almost by an order of magnitude smaller than the result (\ref{ChiralVorticalConductivity}) in the deconfinement phase at the temperature $T = 1.58 \, T_c$. However, one can expect that by virtue of asymptotic freedom in the limit of very high temperatures the chiral vortical conductivity approaches its value (\ref{ChiralVorticalConductivity}) for free fermions.

 On the other hand, in the recent work \cite{Buividovich:13:6} it was demonstrated that anomalous transport phenomena such as Chiral Magnetic Effect are extremely sensitive to finite volume and finite temperature effects. In particular, it was found that lattice measurements of CME with free chiral lattice fermions in the background of a constant magnetic field in a finite volume yield only half of the result which one can obtain from anomaly-based arguments. Preliminary studies have also indicated that the value of the chiral magnetic conductivity is also extremely sensitive to lattice artifacts which lead to the non-conservation of the vector current. Such artifacts are also present in the measurements of \cite{Braguta:13:1}, since the naive discretization of the energy flow vector used in this work is in general not conserved. In view of these considerations, it is reasonable to ask whether the discrepancy between the lattice results of \cite{Braguta:13:1} in the deconfinement phase and the free-fermion result (\ref{ChiralVorticalConductivity}) is due to the lattice or finite-volume artifacts or due to the genuinely strong-coupling nature of the quark-gluon plasma state.

 In these Proceedings we calculate the chiral vortical conductivity for free overlap fermions on the lattice and check the validity of the result (\ref{ChiralVorticalConductivity}). We use the same lattice setup as in \cite{Braguta:13:1} and study the naively discretized energy flow along the direction of the constant axial magnetic field.

 As demonstrated in \cite{Buividovich:13:6}, overlap Dirac operator in the presence of both vector and axial gauge fields $V_{\mu}\lr{x}$ and $A_{\mu}\lr{x}$ should satisfy the following functional equation:
\begin{eqnarray}
\label{LatticeDiracAxialDefiningEquation}
  \frac{\partial}{\partial A_{\mu}\lr{x}} \mathcal{D}_{ov}\lrs{V_{\mu}\lr{x}, A_{\mu}\lr{x}}
  =
  \frac{\partial}{\partial V_{\mu}\lr{x}} \mathcal{D}_{ov}\lrs{V_{\mu}\lr{x}, A_{\mu}\lr{x}}
  \gamma_5 \lr{1 - \mathcal{D}_{ov}\lrs{V_{\mu}\lr{x}, A_{\mu}\lr{x}}} .
\end{eqnarray}
This equation drastically simplifies if one formulates it in terms of the overlap Dirac propagator $\mathcal{D}_{ov}^{-1}$:
\begin{eqnarray}
\label{PropagatorEquation}
  \frac{\partial}{\partial A_{\mu}\lr{x}} \, \mathcal{D}_{ov}^{-1}
  =
  - \mathcal{D}_{ov}^{-1} \, \lr{\frac{\partial}{\partial A_{\mu}\lr{x}} \mathcal{D}_{ov}} \, \mathcal{D}_{ov}^{-1}
  =
  - \mathcal{D}_{ov}^{-1} \, \lr{\frac{\partial}{\partial V_{\mu}\lr{x}} \mathcal{D}_{ov}} \,
  \gamma_5 \lr{1 - \mathcal{D}_{ov}} \,
  \mathcal{D}_{ov}^{-1}
 = \nonumber \\ =
 \mathcal{D}_{ov}^{-1} \, \lr{\frac{\partial}{\partial V_{\mu}\lr{x}} \mathcal{D}_{ov}} \, \mathcal{D}_{ov}^{-1} \gamma_5 = - \frac{\partial}{\partial V_{\mu}\lr{x}} \mathcal{D}_{ov}^{-1} \gamma_5  ,
\end{eqnarray}
where in the last line we have used the Ginsparg-Wilson relation and the arguments of $\mathcal{D}_{ov}$ were not shown for the sake of brevity. We have thus arrived at the linear equation, which has a straightforward solution in terms of the chiral projectors $\mathcal{P}_{\pm} = \frac{1 \pm \gamma_5}{2}$:
\begin{eqnarray}
\label{PropagatorSolution}
 \mathcal{D}_{ov}^{-1}\lrs{V_{\mu}, A_{\mu}} =
 \mathcal{P}_{+} \mathcal{D}_{ov}^{-1}\lrs{V_{\mu} + A_{\mu}} \mathcal{P}_{-}
 +
 \mathcal{P}_{-} \mathcal{D}_{ov}^{-1}\lrs{V_{\mu} - A_{\mu}} \mathcal{P}_{+}
 + 1/2 ,
\end{eqnarray}
where the propagators $\mathcal{D}_{ov}^{-1}\lrs{V_{\mu} \pm A_{\mu}}$ are obtained from the overlap Dirac operators $\mathcal{D}_{ov}\lrs{V_{\mu} \pm A_{\mu}}$ in the background of the vector gauge fields $V_{\mu}\lr{x} \pm A_{\mu}\lr{x}$. The latter are coupled to fermions in a standard way by including the corresponding link factors into finite difference operators entering the Dirac-Wilson operator which is used to define the overlap Dirac operator \cite{Neuberger:98:1}. The integration constant $1/2$ in (\ref{PropagatorSolution}) is fixed by the requirement that at $A_{\mu}\lr{x}=0$ one obtains the conventional overlap propagator in the background of the vector field $V_{\mu}\lr{x}$ \cite{Neuberger:98:1}. A similar construction for the overlap Dirac propagator in the presence of an axial gauge field was also used in \cite{Braguta:13:1}, however, an integration constant of $1/2$ was not taken into account. It is indeed irrelevant for the Chiral Vortical Effect. However, this constant in fact incorporates the effect of axial anomaly on the lattice \cite{Neuberger:98:1} and can be extremely important for other anomalous transport phenomena such as Chiral Magnetic Effect or Chiral Separation Effect.

 In order to implement the constant axial magnetic field on the lattice, we use the configuration of $A_{\mu}\lr{x}$ which yields the constant lattice field strength $F_{xy} = -F_{yx} = A_{x}\lr{x} + A_{y}\lr{x + \hat{e}_x} - A_{x}\lr{x + \hat{e}_y} - A_{y}\lr{x}$ in (\ref{PropagatorSolution}) and set $V_{\mu}\lr{x} \equiv 0$. Due to flux quantization, possible values of field strength are $F_{xy} = 2 \pi N_5/L^2$, where $L$ is the spatial lattice size and $N_5$ is an integer. Chiral and baryon chemical potentials are set to zero. Overlap Dirac operator is calculated and inverted using \texttt{LAPACK}. The energy flow along the axial magnetic field $J_E \equiv T_{0 z}$ (\ref{EnergyMomentumTensor}) is then calculated as in \cite{Braguta:13:1}.

\begin{figure}[htpb]
  \centering
  \includegraphics[width=6.5cm, angle=-90]{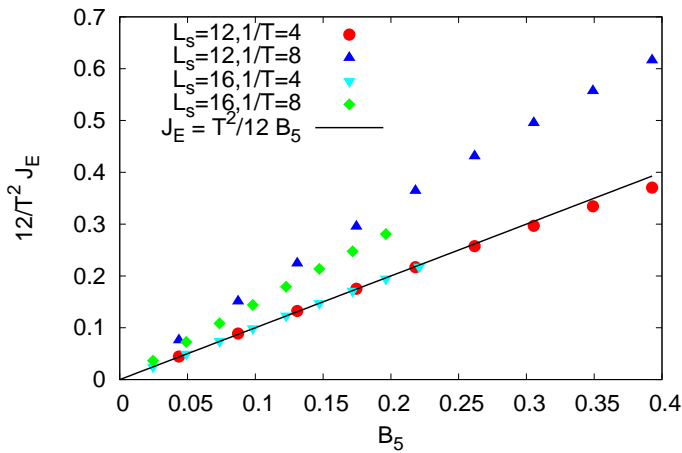}\\
  \caption{Energy flow (\ref{EnergyMomentumTensor}) on the lattice as a function of axial magnetic field $B_5$ at different spatial lattice sizes and at different temperatures. Solid line corresponds to the continuum result (\ref{CVEClassical}), (\ref{ChiralVorticalConductivity}).}
  \label{fig:cve_vs_NB5}
\end{figure}

 On Figure \ref{fig:cve_vs_NB5} we plot the energy flow $J_E$ along the direction of the axial magnetic field as a function of field strength for different temperatures and different spatial lattice sizes. We assume that lattice spacing is equal to unity. One can see that in accordance with (\ref{CVEClassical}) the energy flow is almost a linear function of $B_5$ with a slight tendency for saturation at large values of $B_5$.

\begin{figure*}[htpb]
  \centering
  \includegraphics[width=5.5cm, angle=-90]{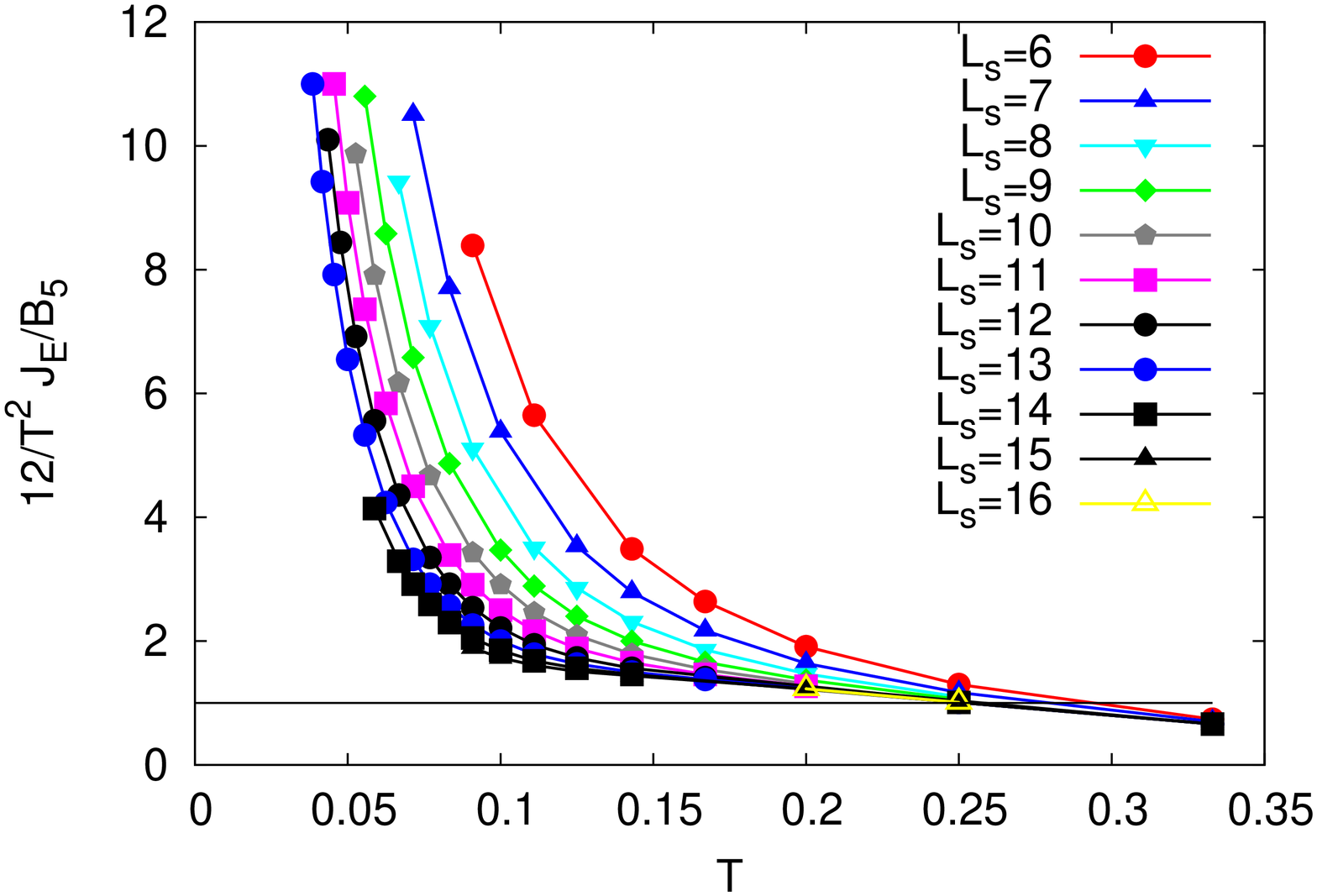}
  \includegraphics[width=5.5cm, angle=-90]{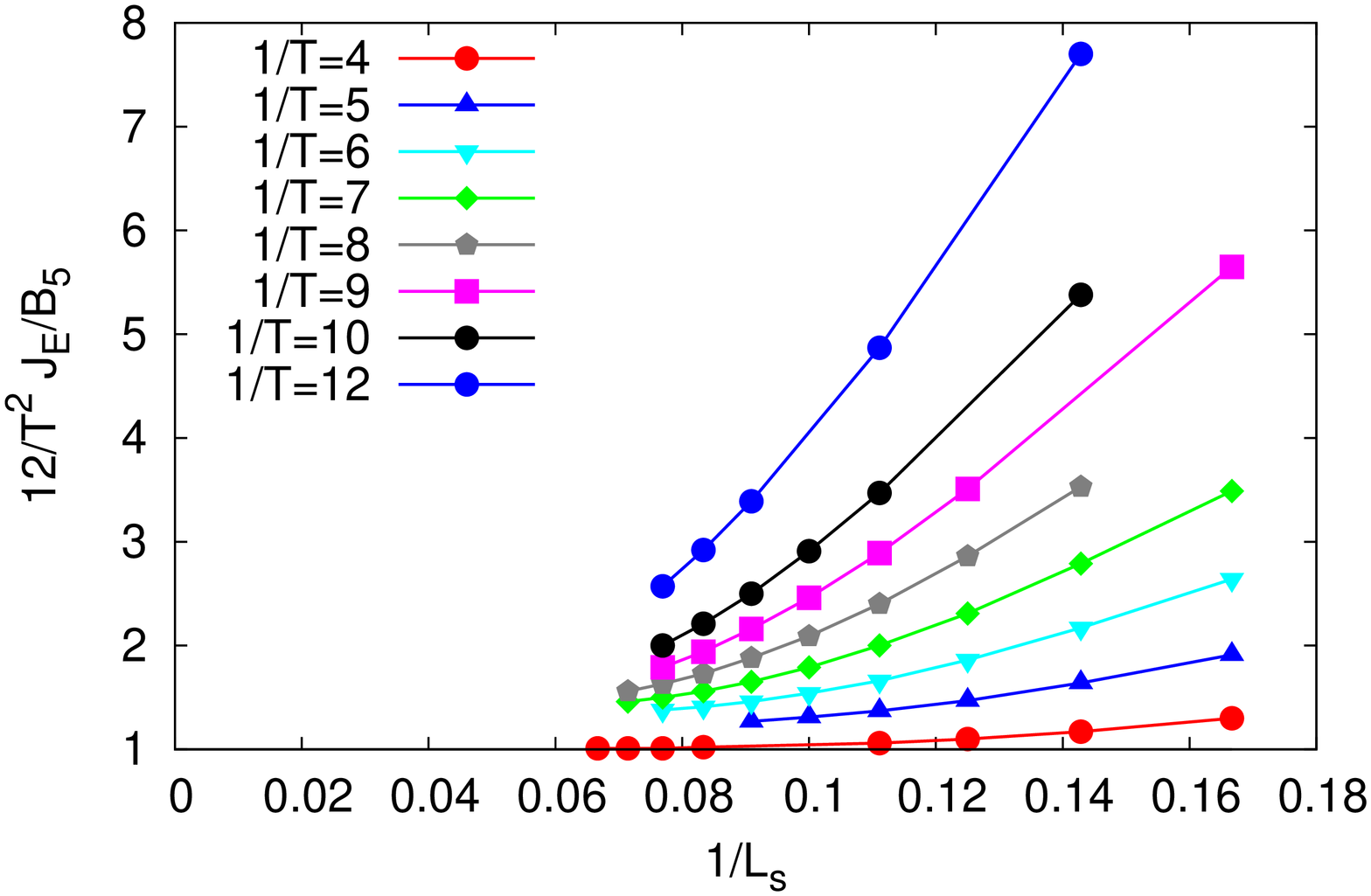}\\
  \caption{Finite volume scaling of the chiral vortical conductivity calculated with free overlap lattice fermions. On the left: chiral vortical conductivity as a function of temperature at different spatial lattice sizes. On the right: chiral vortical conductivity as a function of inverse lattice size at different fixed temperatures.}
  \label{fig:cve_finite_volume}
\end{figure*}

 In order to extract the value of the chiral vortical conductivity, we perform the linear fits of the energy flow as a function of $B_5$, using data points with three smallest values of $B_5$. The results are illustrated on Fig. \ref{fig:cve_finite_volume} for different temperatures and different lattice sizes. An immediate conclusion is that the finite-volume effects are very strong, several times stronger than the continuum, infinite-volume result (\ref{ChiralVorticalConductivity}). However, from the right plot on Fig. \ref{fig:cve_finite_volume} one can also see that if one increases the lattice size at fixed temperature, the chiral vortical conductivity indeed tends to the continuum result (\ref{ChiralVorticalConductivity}). The smaller is the temperature, the larger should be the lattice size.

 We conclude that the continuum result (\ref{ChiralVorticalConductivity}) is indeed reproduced for free overlap lattice fermions, however, finite-volume effects are very strong and one should perform a very careful extrapolation to the thermodynamic limit.

\section*{Acknowledgements}

 I would like to dedicate this work to the memory of my Teacher, Prof.~Dr. M.~I.~Polikarpov. I am grateful to Dr. V. Braguta, Dr. M. Chernodub, Dr. K. Landsteiner and Dr. M. Ulybyshev for interesting discussions. This work was supported by the S. Kowalewskaja award from the Alexander von Humboldt foundation.

\section*{References}


\begin{thebibliography}{10}
\expandafter\ifx\csname url\endcsname\relax
  \def\url#1{{\tt #1}}\fi
\expandafter\ifx\csname urlprefix\endcsname\relax\def\urlprefix{URL }\fi
\providecommand{\eprint}[2][]{\url{#2}}

\bibitem{Kharzeev:12:1}
Kharzeev D~E, Landsteiner K, Schmitt A and Yee H 2012 Strongly interacting
  matter in magnetic fields: an overview in Lect. Notes Phys. "Strongly
  interacting matter in magnetic fields" (Springer)
  \href{http://arxiv.org/abs/1211.6245}{ArXiv:1211.6245}

\bibitem{Hongo:13:1}
Hongo M, Hirono Y and Hirano T 2013 First numerical simulations of anomalous
  hydrodynamics \href{http://arxiv.org/abs/1309.2823}{ArXiv:1309.2823}

\bibitem{Erdmenger:08:1}
Erdmenger J, Haack M, Kaminski M and Yarom A 2009 {\em JHEP\/} {\bf 0901} 055
  \href{http://arxiv.org/abs/0809.2488}{ArXiv:0809.2488}

\bibitem{Surowka:08:1}
Banerjee N, Bhattacharya J, Bhattacharyya S, Dutta S, Loganayagam R and
  Sur\'{o}wka P 2011 {\em JHEP\/} {\bf 1101} 094
  \href{http://arxiv.org/abs/0809.2596}{ArXiv:0809.2596}

\bibitem{Landsteiner:11:1}
Landsteiner K, Megias E and {Pena-Benitez} F 2011 {\em Phys. Rev. Lett.\/} {\bf
  107} 021601 \href{http://arxiv.org/abs/1103.5006}{ArXiv:1103.5006}

\bibitem{Son:12:1}
Golkar S and Son D~T 2012 (Non)-renormalization of the chiral vortical effect
  coefficient \href{http://arxiv.org/abs/1207.5806}{ArXiv:1207.5806}

\bibitem{Ren:12:1}
Hou D, Liu H and Ren H 2012 {\em Phys. Rev. D\/} {\bf 86} 121703
  \href{http://arxiv.org/abs/1210.0969}{ArXiv:1210.0969}

\bibitem{Landsteiner:11:2}
Amado I, Landsteiner K and Pena-Benitez F 2011 {\em JHEP\/} {\bf 05} 081
  \href{http://arxiv.org/abs/1102.4577}{ArXiv:1102.4577}

\bibitem{Yamamoto:13:1}
Yamamoto A and Hirono Y 2013 {\em Phys. Rev. Lett.\/} {\bf 111} 081601
  \href{http://arxiv.org/abs/1303.6292}{ArXiv:1303.6292}

\bibitem{Yamamoto:11:1}
Yamamoto A 2011 {\em Phys. Rev. Lett.\/} {\bf 107} 031601
  \href{http://arxiv.org/abs/1105.0385}{ArXiv:1105.0385}

\bibitem{Braguta:13:1}
Braguta V, Chernodub M~N, Landsteiner K, Polikarpov M~I and Ulybyshev M~V 2013
  Numerical evidence of the axial magnetic effect
  \href{http://arxiv.org/abs/1303.6266}{ArXiv:1303.6266}

\bibitem{Buividovich:13:6}
Buividovich P~V 2013 {\em PoS\/} {\bf Lattice2013} 179
  \href{http://arxiv.org/abs/1309.2850}{ArXiv:1309.2850}

\bibitem{Neuberger:98:1}
Neuberger H 1998 {\em Phys. Lett. B\/} {\bf 417} 141
  \href{http://arxiv.org/abs/hep-lat/9707022}{ArXiv:hep-lat/9707022}
\end{thebibliography}

\providecommand{\newblock}{}

\end{document}